\documentclass[prl,twocolumn,aps,showpacs]{revtex4}

\usepackage{epsfig}
\usepackage{graphicx}

\begin{document}

\title{Temperature and coupling dependence of the universal contact intensity \\ for an ultracold Fermi gas}

\author{F. Palestini, A. Perali, P. Pieri, and G. C. Strinati}

\affiliation{Dipartimento di Fisica, Universit\`{a} di Camerino, I-62032 Camerino, Italy}

\begin{abstract}
Physical properties of an ultracold Fermi gas in the temperature-coupling phase diagram can be characterized by the contact intensity
$\mathrm{C}$, which enters the pair-correlation function at short distances and describes how the two-body problem merges into its surrounding.
We show that the local order established by pairing fluctuations about the
critical temperature $T_{c}$ of the superfluid transition considerably enhances the contact $\mathrm{C}$ in a temperature range where pseudogap phenomena are maximal.
Our \emph{ab initio} results for $\mathrm{C}$ in a trap compare well with recently available experimental data over a wide coupling range. 
An analysis is also provided for the effects of trap averaging on $\mathrm{C}$.
\end{abstract}

\maketitle

The ``contact'' $\mathrm{C}$, introduced by Tan \cite{Tan-2008-I,Tan-2008-II} to characterize the merging of two-body into many-body physics in systems like ultracold Fermi gases with a short-range inter-particle interaction, has attracted much interest lately \cite{Braaten-2008,Castin-2009,Leggett-2009,Combescot-2009,Castin-2010}.
This is especially relevant in the context of the BCS-BEC crossover, whereby a smooth evolution occurs 
jointly for the two- and many-body physics, from the presence of Cooper pairs with an underlying Fermi surface in the BCS limit, to the formation of molecular bosons with a residual interaction in the BEC limit.

Recently, the contact $\mathrm{C}$ was measured in an ultracold gas of trapped fermionic ($^{40}\mathrm{K}$) atoms
\cite{Jin-2010}, from the large-momentum tail of the momentum distribution as well as from the high-frequency tail of the radio-frequency signal following an earlier suggestion \cite{PPS-2009}.
These measurements were done from about unitarity (where the scattering length $a_{F}$ of the Fano-Feshbach resonance diverges) to deep inside the BCS region (more precisely, in the coupling range $- 2.5 \lesssim (k_{F} a_{F})^{-1} \lesssim + 0.5$ where $k_{F} = (2 m E_{F})^{1/2}$ is the Fermi wave vector expressed in terms of the Fermi energy $E_{F} = \omega_{0} (3 N)^{1/3}$, $m$ being the atom mass, $N$ the total number of atoms, and $\omega_{0}$ the average trap frequency \cite{footnote-1}), and in a temperature range about the critical temperature $T_{c}$.

Several questions concerning the contact $\mathrm{C}$ remain open. 
They hinge on the recent experimental results of Ref.\cite{Jin-2010} (like the temperature and coupling dependence of $\mathrm{C}$, and the effect that trap averaging has on the values of $\mathrm{C}$), as well as on more theoretical issues. 
These include the identification of the (approximate) spatial boundary between short- and medium-range physics that can be associated with the contact $\mathrm{C}$, the effects that improved theoretical approaches have on the values of $\mathrm{C}$, and the interconnection with the presence of a pseudogap in the single-particle excitation spectrum. 

In this Letter, we address these questions and calculate the contact $\mathrm{C}$ using a t-matrix approach \cite{PPSC-2002} that proved successful in comparison with ARPES-like data for ultracold Fermi atoms \cite{JILA-Cam-2009}, and also using a nontrivial extension of this theory \cite{PS-2005} which takes into account the residual interaction among composite bosons.
This is to verify to what an extent improvements on the description of the medium-range physics (over and above the results of the t-matrix) influence the values of $\mathrm{C}$ in different coupling and temperature ranges \cite{Baym-2009}.

We shall, specifically, be concerned with the temperature dependence of $\mathrm{C}$ over an \emph{extended\/} temperature range up to (several times) the Fermi temperature $T_{F}$, to determine how the value of $\mathrm{C}$ is affected by the pseudogap physics extending above $T_{c}$ in the unitary ($- 1 \lesssim (k_{F} a_{F})^{-1} \lesssim + 1$) regime, and to address the related question of how trap averaging influences the value of $\mathrm{C}$ with respect to that of a homogeneous system with the same nominal temperature and coupling.

The contact $\mathrm{C}$ was originally introduced to account for the large wave-vector behavior of the fermionic distribution $n(k)$ (for spin component) of the homogeneous system, such that $n(k) \approx \mathrm{C_{h}} \, k^{-4}$. 
Here, the suffix ``h'' stands for homogeneous, $k = |\mathbf{k}|$ is in units of $k_{F} = (3 \pi^{2} n)^{1/3}$ where $n$ is the total particle density, and  $n(k)$ is normalized such that $\int \! \frac{d\mathbf{k}}{(2 \pi)^{3}} \, n(k) = 1/2$.
Alternatively, $\mathrm{C_{h}}$ can be extracted from the high-frequency tail of the radio-frequency (RF) spectrum $I_{RF}(\omega)$ per unit volume, so that $I_{RF}(\omega) \approx (\mathrm{C_{h}}/ 2^{3/2} \pi^{2}) \, \omega^{-3/2}$.
Here, the frequency $\omega$ is in units of $E_{F}$ and the RF spectrum is normalized such that
$ \int _{- \infty}^{+ \infty} \! d\omega \, I_{RF}(\omega) = 1/2$.
The above asymptotic form of $I_{RF}(\omega)$ holds provided final-state effects can be neglected \cite{PPS-2009,Randeria-2009,Braaten-2010}.

Similar asymptotic behaviors can be obtained for the trapped system. 
Preserving the above normalization as for the homogeneous system, we write within a local-density approximation:
\begin{equation}
n(k) \, = \, \int \! d\mathbf{r} \, n(k;\mathbf{r}) \, \approx \, \frac{\mathrm{C_{t}}}{k^{4}}                           \label{n-asymptotic-trap}
\end{equation}

\noindent
where 
\begin{equation}
\mathrm{C_{t}} \, = \, \frac{8}{\pi^{2}} \, \int \! d\mathbf{r} \, \frac{[3 \pi^{2} \, n(\mathbf{r})]^{4/3}}{k_{F}^{4}} \,\, \mathrm{C_{h}}(\mathbf{r})   \,\, .                                  
                                                                                                                                                                    \label{C-trap-vs-homogeneous}
\end{equation}

\noindent 
Here, the suffix ``t'' and $k_{F}$ refer to the trapped system, the spatial position $\mathbf{r}$ is in units of the Thomas-Fermi radius $R_{F} = [2 E_{F}/(m \omega_{0}^{2})]^{1/2}$, and $n(\mathbf{r})$ and $\mathrm{C_{h}}(\mathbf{r})$ are the density and contact \emph{locally\/} in the trap.
By a similar token, the large-$\omega$ behavior of the total RF spectrum of the trapped system reduces to:
\begin{equation}
I_{RF}(\omega) \, = \, \int \! d\mathbf{r} \, I_{RF}(\omega;\mathbf{r}) \, \approx \, \frac{\mathrm{C_{t}}}{2^{3/2} \pi^{2} \, \omega^{3/2}}                           
                                                                                                                                                                          \label{RF-asymptotic-trap}
\end{equation}

\noindent
with the overall normalization of the homogeneous case.

This local-density analysis shows that, by adding contributions of different shells with a weight proportional to $n(\mathbf{r})^{4/3}$, the trap modifies the value $\mathrm{C_{h}}$ for a homogeneous system at the same nominal coupling $(k_{F} a_{F})^{-1}$ and (relative) temperature $T/T_{F}$.
In this respect, the temperature dependence of $\mathrm{C_{h}}$ provides an important information, because different shells are at different relative temperature with respect to the local value of $T_{F}$.

\begin{figure}[t]
\begin{center}
\includegraphics[angle=0,width=7cm]{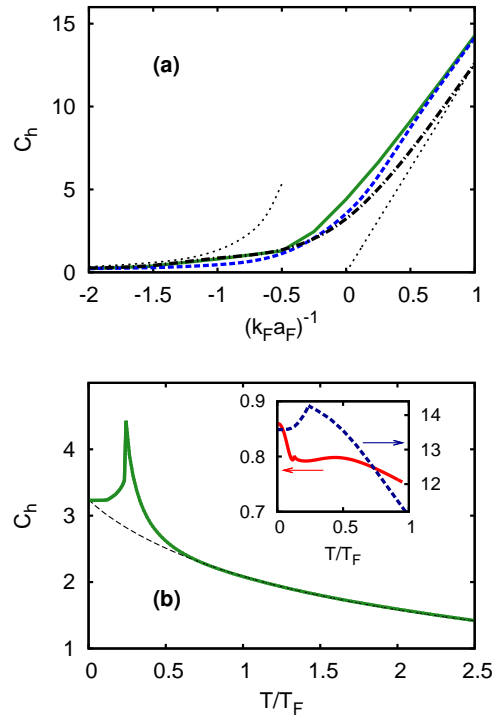}
\caption{ The contact $\mathrm{C_{h}}$ for the homogeneous case: (a) At $T_{c}$ vs the coupling $(k_{F} a_{F})^{-1}$, obtained within the t-matrix approximation (full line) and its improved Popov version (dashed line).
The leading approximations in weak and strong coupling (light-dotted lines) as well as the $T=0$ result within the t-matrix approximation (dashed-dotted line) are also reported.
(b) At unitarity vs the temperature $T$ (in units of the Fermi temperature $T_{F}$), obtained within the t-matrix (full line). 
The high-temperature approximation to this curve is also reported as a reference (dashed line) and extrapolated to $T=0$. 
[See the text for the meaning of the inset.]}
\label{fig1}
\end{center}
\end{figure}

We begin our analysis with the homogeneous case at $T_{c}$ and show in Fig.~\ref{fig1}(a) the coupling dependence of $\mathrm{C_{h}}$, when $n(k)$ is obtained within the t-matrix approximation of Ref.\cite{PPSC-2002} (full line) and its improved Popov version of Ref.\cite{PS-2005} (dashed line). 
The calculations are done at the value of $T_{c}$ of the respective theories.
The plot also shows (light-dotted lines) the leading (low-temperature) approximations for $\mathrm{C_{h}}$ obtained in the BCS weak-coupling and BEC strong-coupling limits, which are given by the expressions $(4/3) k_{F}^{2} a_{F}^{2}$ and $4 \pi /(k_{F} a_{F})$, respectively,
with the crossover region $- 1 \lesssim (k_{F} a_{F})^{-1} \lesssim + 1$ marking the change between these two limiting behaviors.

Note how the difference between the full and dashed lines in Fig.~\ref{fig1}(a), which originates from the activation of the residual interaction among the composite bosons, is appreciable only close to unitarity where particles correlate with each other within the inter-particle spacing $k_{F}^{-1}$.
The smallness of this difference resulting from our calculations confirms the validity of the t-matrix approximation for the contact 
$\mathrm{C_{h}}$ and for the high-energy scale to which $\mathrm{C_{h}}$ is associated (see below).
Yet, this difference is relevant for the physical interpretation of the contact $\mathrm{C_{h}}$ as characterizing the effects of medium-range (many-body) physics over and above the short-range (two-body) physics.
This interpretation is also consistent with rewriting $\mathrm{C_{h}} = (3 \pi^{2} /4) (\Delta_{\infty} /E_{F})^{2}$ in terms of the high-energy scale $\Delta_{\infty}$ introduced in Ref.\cite{PPS-2009}, such that $\Delta_{\infty} = 2 \pi |a_{F}| n / m$ embodies in weak coupling the effects of surrounding particles through a mean-field shift \cite{footnote-2}, while $\Delta_{\infty}^{2} = 4 \pi n /(m^{2} a_{F})$ reflects a standard relation in strong coupling \cite{PS-2003} between the density and the gap parameter within BCS theory.
For comparison, Fig.~\ref{fig1}(a) also reports the coupling dependence of $\mathrm{C_{h}}$ at zero temperature (dashed-dotted line) within the t-matrix approximation, to which the above approximate expressions in the BCS and BEC limits converge.

Figure \ref{fig1}(b) shows the temperature dependence of $\mathrm{C_{h}}$ at unitarity over an extended temperature range within the t-matrix approximation.
The rather slow decay of $\mathrm{C_{h}}$ at high temperature is consistent with the expectation that $\mathrm{C_{h}}$ is not related to long-range order.
The temperature behavior of $\mathrm{C_{h}}$ steepens up at low temperature when entering the pseudogap region for $T/T_{F} \lesssim 0.5$ \cite{JILA-Cam-2009}, thus evidencing the emergence of a local (medium-range) order which is also responsible for the pseudogap.
Such an enhancement of the value of $\mathrm{C_{h}}$ appears most evident when the calculation is continued below $T_{c}$ \cite{PPS-2004}, with the result that a cusp appears in $\mathrm{C_{h}}$ at $T_{c}$ where the effect of the pseudogap is maximum.
To emphasize this enhancement, we have indicated in Fig.~\ref{fig1}(b) the extrapolation of the high-temperature behavior of $\mathrm{C_{h}}$ (dashed line) down to $T=0$, on top of which the contribution associated with the pseudogap region about $T_{c}$ appears evident.
Our value (=3.23) for $\mathrm{C_{h}}$ at $T=0$ compares well with that (=3.40) extracted from the Monte-Carlo calculations of 
Ref.\cite{CGS-2006}.
For completeness, the inset of Fig.~\ref{fig1}(b) reports the temperature dependence of $\mathrm{C_{h}}$ for $(k_{F} a_{F})^{-1} = -1.0$ (full line/left scale) and $(k_{F} a_{F})^{-1} = +1.0$ (dashed line/right scale).
The maximum at $T/T_{F} \simeq 0.5$ in the weak-coupling curve is consistent with the Fermi-liquid behavior discussed in Ref.\cite{Baym-2009}.                                                                                                                                                             

\begin{figure}[t]
\begin{center}
\includegraphics[angle=0,width=8cm]{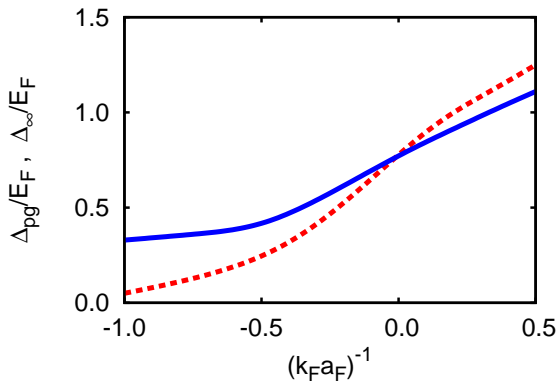}
\caption{The high-energy scale $\Delta_{\infty}$ (full line) and the low-energy scale $\Delta_{\mathrm{pg}}$ (dashed line) are reported 
(in units of $E_{F}$) vs the coupling $(k_{F} a_{F})^{-1}$ at $T_{c}$.}
\label{fig2}
\end{center}
\end{figure}

The temperature dependence of $\mathrm{C_{h}}$ is here reported for the first time and deserves further comments.
On physical grounds, the enhancement of $\mathrm{C_{h}}$ when entering the fluctuative region from above $T_{c}$ is due to the strengthening of local pairing correlations, still in the absence of long-range order.
In the present approach, pairing correlations are embodied by the pair-fluctuation propagator in the particle-particle channel, whose
wave-vector and frequency structures give rise to a characteristic low-energy scale $\Delta_{\mathrm{pg}}$ in the single-particle excitations \cite{PPSC-2002}, which is referred to as the pseudogap.
The contact $\mathrm{C_{h}}$, through its alternative definition in terms of the high-energy scale $\Delta_{\infty}$ that was previously mentioned, is also related to a wave-vector and frequency averaging of the very same structures in the pair-fluctuation propagator 
\cite{PPS-2009}.
The inter-dependence between the two energy scales $\Delta_{\infty}$ and $\Delta_{\mathrm{pg}}$ can then be explicitly  appreciated in Fig.~\ref{fig2}, where they are shown at $T_{c}$ vs the coupling $(k_{F} a_{F})^{-1}$.  
The two energy scales do not fully relate to each other in weak coupling where the contact is dominated by the mean-field interaction, but become very close in value at unitarity where strong local-pairing correlations dominate both thermodynamic ($\Delta_{\infty}$) and dynamic ($\Delta_{\mathrm{pg}}$) quantities.                                                

\begin{figure}[t]
\begin{center}
\includegraphics[angle=0,width=8cm]{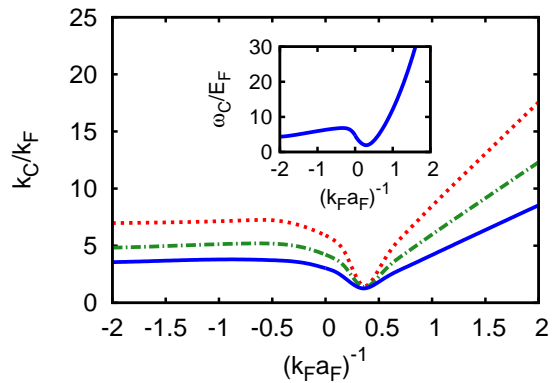}
\caption{The characteristic values of $k_{\mathrm{C}}$ (in units of $k_{F}$), at which the asymptotic power-law behavior $n(k) \approx \mathrm{C_{h}} \, k^{-4}$ is reached within $2.5 \%$ (dashed line), $5 \%$ (dashed-dotted line), and $10 \%$ (full line) accuracy, 
are shown at $T_{c}$ vs $(k_{F} a_{F})^{-1}$.
The inset shows, correspondingly, the characteristic value $\omega_{C}$ (in units of $E_{F}$), at which the asymptotic power-law behavior $I_{RF}(\omega) \approx (\mathrm{C_{h}}/ 2^{3/2} \pi^{2}) \, \omega^{-3/2}$ of the RF spectrum is reached within
$10 \%$ accuracy.}
\label{fig3}
\end{center}
\end{figure}

Although the values of $\mathrm{C_{h}}$ are extracted from the asymptotic (scale-free) power law $n(k) \approx \mathrm{C_{h}} \, k^{-4}$, a characteristic value $k_{\mathrm{C}}$ can nevertheless be identified at which $n(k)$ has reached $\mathrm{C_{h}} \, k^{-4}$ within, say, a few percent accuracy.
The length scale corresponding to $k_{\mathrm{C}}^{-1}$ is approximately the spatial range at which two-body physics interfaces with medium-range physics.
This range is expected to be about the size of the composite bosons in strong coupling and the inter-particle spacing in weak coupling.
Figure \ref{fig3} shows the values of $k_{\mathrm{C}}$ extracted in this way at $T_{c}$ vs $(k_{F} a_{F})^{-1}$ within the t-matrix approximation, for the three distinct values $(2.5,5,10)\%$ of the above percent accuracy.
In extreme weak coupling, our numerical values recover the limiting ones obtained for a dilute Fermi gas \cite{Belyakov-1961}.
Note how $k_{\mathrm{C}}$ reaches a pronounced minimum for the coupling at which the chemical potential vanishes, away from which $k_{\mathrm{C}}$ increases most markedly on the strong-coupling side as expected from the increasing spatial localization of composite bosons.

In addition, the inset of Fig.~\ref{fig3} shows the value of $\omega_{\mathrm{C}}$ at $T_{c}$ vs $(k_{F} a_{F})^{-1}$, extracted within a $10 \%$ accuracy from the large-$\omega$ behavior of the RF spectrum.
(Recall that the $\omega^{-3/2}$ tail of the RF spectrum originates from the short-range behavior of the 2-body wave function \cite{Chin-Julienne-2005,PPS-2008} in the absence of final-state effects \cite{footnote-3}.)
Comparison with $k_{\mathrm{C}}$ with the same accuracy yields the relation $2 \omega_{C} = k_{C}^{2}$ that holds approximately for all couplings.

\begin{figure}[t]
\begin{center}
\includegraphics[angle=0,width=8cm]{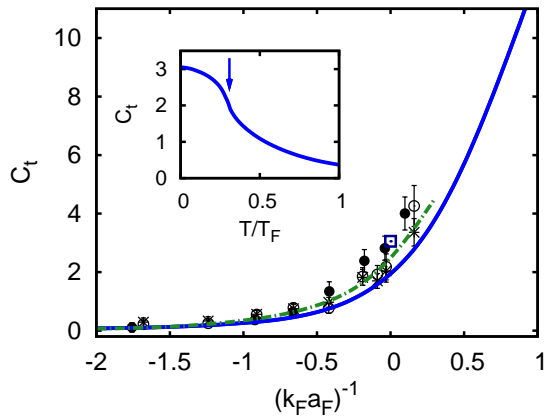}
\caption{The contact $\mathrm{C_{t}}$ obtained within the t-matrix approximation for the trapped case (full line) is shown  
at $T_{c}$ vs the coupling $(k_{F} a_{F})^{-1}$, and compared with the experimental values of Ref.\cite{Jin-2010} (filled and empty circles, stars).
The inset shows, correspondingly, $\mathrm{C_{t}}$ at unitarity vs $T/T_{F}$.}
\label{fig4}
\end{center}
\end{figure}

The values of the contact $\mathrm{C_{t}}$ obtained from Eq.(\ref{n-asymptotic-trap}), by adding the asymptotic contributions from all shells in the trap within the t-matrix approximation, are reported in Fig.~\ref{fig4} at $T_{c}$ vs $(k_{F} a_{F})^{-1}$ (full line).
They are compared with the values (filled and empty circles, stars) obtained experimentally in Ref.\cite{Jin-2010} 
through alternative procedures.
The theoretical value obtained at unitarity for $T=0$ is also reported for comparison (empty square).
Due to difficulties in extracting the asymptotic behavior of $n(k)$ from
experimental data, the figure also shows the theoretical values of
$\mathrm{C_{t}}$ (dashed-dotted line) obtained upon averaging $k^{4} n(k)$
over the interval $k_{\mathrm{min}} \le k \le k_{\mathrm{max}}$. 
This follows the procedure used to extract the experimental values of $\mathrm{C_{t}}$, for which $k_{\mathrm{min}} = 1.55$ when $(k_{F} a_{F})^{-1} < -0.5$ and $k_{\mathrm{min}} = 1.85$ when $-0.5 < (k_{F} a_{F})^{-1}$, while $k_{\mathrm{max}} = 2.5$ \cite{Jin-2010}.

It is worth noting that the data of Ref.\cite{Jin-2010} which are reported in Fig.~\ref{fig4}, have been originally compared with a theoretical curve from Ref.\cite{Castin-2009}, where the behavior of $\mathrm{C_{t}}$ at $T=0$ was obtained across the unitary regime by interpolating known results in the BCS and BEC limits.
That interpolation has to be contrasted with the completely \emph{ab initio} theoretical calculation at $T_{c}$ reported by the full line in Fig.~\ref{fig4}, which spans a wide coupling range just across the unitary regime.

Finally, the inset of Fig.~\ref{fig4} displays the temperature dependence of $\mathrm{C_{t}}$ for the trapped case at unitarity within the t-matrix approximation (full line), with the vertical arrow indicating the corresponding value of $T_{c}$.
Contrary to the result for the homogeneous case of Fig.~\ref{fig1}(b), no cusp now appears in $\mathrm{C_{t}}$ at $T_{c}$ for the trapped case.
An expanded discussion about the effects that trap averaging has on the values of $\mathrm{C}$ is provided in \cite{SM-PRL}.

In conclusion, we have presented a detailed analysis of the contact intensity $\mathrm{C}$ for a Fermi gas over the temperature-coupling phase diagram.
The values of $\mathrm{C}$ have been obtained from the large-$k$ behavior of the momentum distribution $n(k)$ as well as from the large-$\omega$ tail of the RF spectrum $I_{RF}(\omega)$, both for the homogeneous and trapped case.
For the latter case, good agreement is obtained with recent experimental data. 
The effects of pairing fluctuations on $\mathrm{C}$ have been determined by the t-matrix approximation and its improved Popov version that takes into account the interaction among composite bosons.
We have found that the values of $\mathrm{C}$ are strongly affected by the emergence of a pseudogap in the single-particle excitations about $T_{c}$. 

\acknowledgments
Discussions with E. Cornell, D. Jin, and J. T. Stewart are gratefully
acknowledged. This work was partially supported by the Italian MIUR under
Contract PRIN-2007 ``Ultracold Atoms and Novel Quantum Phases''.



\newpage 

\begin{center}
{\bf SUPPLEMENTAL MATERIAL: ``TEMPERATURE AND COUPLING DEPENDENCE OF THE
  UNIVERSAL CONTACT INTENSITY FOR AN ULTRACOLD FERMI GAS''}
\end{center}

{\em We discuss the connection between the values of the contact intensity for
  the whole trap and for a homogeneous system.}

It is interesting to compare the values of $\mathrm{C_{t}}$ obtained for the whole trap with its approximation obtained from Eq.(2) of the main text, whereby one identifies the shell at $r_{\mathrm{max}}$ corresponding to the maximum of the 
\emph{radial\/} weight function $(32/\pi) \, r ^{2} \, [3 \pi^{2} \, n(r)]^{4/3}/k_{F}^{4}$, and then takes $\mathrm{C_{h}}(r_{\mathrm{max}})$ therein outside the integral.
The result of this procedure at $T_{c}$ is reported vs the coupling $(k_{F} a_{F})^{-1}$ in Fig.~5(a), where the inset shows an example of the shape of the radial weight function (full line/right scale) and of $\mathrm{C_{h}}(r)$ 
(dashed line/left scale) at unitarity.
The good agreement, which results between the calculation for the trap (full line) and the approximation that selects the contribution of the most important shell (dashed line), shows to what an extent the results of $\mathrm{C_{t}}$ for the whole trap can be used, 
together with knowledge of the density profiles, to extract the values of $\mathrm{C_{h}}$ for the homogeneous case.

The same procedure can be applied to interpret the temperature dependence of $\mathrm{C_{t}}$, reported in the inset of Fig.~4 of the main text and reproduced here in Fig.~5(b) for convenience (full line).
In particular, it is interesting to understand how the trap averaging washes out the peak about $T_c$ obtained for the homogeneous case (see Fig.~1(b) of the main text).

To appreciate this effect, we have reported in the inset of Fig.~5(b) the temperature dependence at unitarity of the integral of the radial weight function (full line/right scale) and of $\mathrm{C_{h}}(r_{\mathrm{max}})$ 
(dashed line/left scale) from above to below $T_{c}$.
While $\mathrm{C_{h}}(r_{\mathrm{max}})$ retains the characteristic cusp feature of the homogeneous case (with a maximum at $T = 0.8 T_{c}$), the steady increase of the integrated weight for decreasing temperature more than compensates for the decrease 
of $\mathrm{C_{h}}(r_{\mathrm{max}})$ when $T < 0.8 T_{c}$, thus masking eventually the cusp feature in the integrated quantity.

The same approximate procedure, that resulted in the dashed line of Fig.~5(a), can be applied to reproduce the temperature dependence of $\mathrm{C_{t}}$ at unitarity \emph{above\/} $T_{c}$, because in this case the two functions in the integral of Eq.(2) of the 
main text have a smooth behavior similar to that shown in the inset of Fig.~5(a).
At given $T$ \emph{below\/} $T_{c}$, however, the cusp present in $\mathrm{C_{h}}(r)$ requires us to split it as the sum of a smooth background $\mathrm{C_{h}^{(b)}}(r)$ and of a peaked contribution $\mathrm{C_{h}^{(p)}}(r)$, yielding approximately:
\begin{eqnarray}
\mathrm{C_{t}} \simeq \mathrm{C_{h}^{(b)}}(r_{\mathrm{max}}) \, \frac{32}{\pi} \, \int_{0}^{\infty} \! dr \, r^{2} \, \frac{[3 \pi^{2} \, n(r)]^{4/3}}{k_{F}^{4}} \nonumber \\+ \,  \frac{32}{\pi} \, \int_{0}^{\infty} \! dr \, r^{2} \, \frac{[3 \pi^{2} \,
 n(r)]^{4/3}}{k_{F}^{4}}  \, \mathrm{C_{h}^{(p)}}(r) \, .                                  
                                                                                                                                                 \label{C-trap-vs-homogeneous-approximate}
\end{eqnarray}

\noindent
These approximate results, from above to below $T_{c}$, are shown by the dashed line in Fig.~5(b) (to which the second term on the right-hand side of Eq.(\ref{C-trap-vs-homogeneous-approximate}) below $T_{c}$ gives at most a $15 \%$ contribution).

\begin{figure}[b]
\begin{center}
\vspace{-1cm}
\includegraphics[angle=0,width=8.5cm]{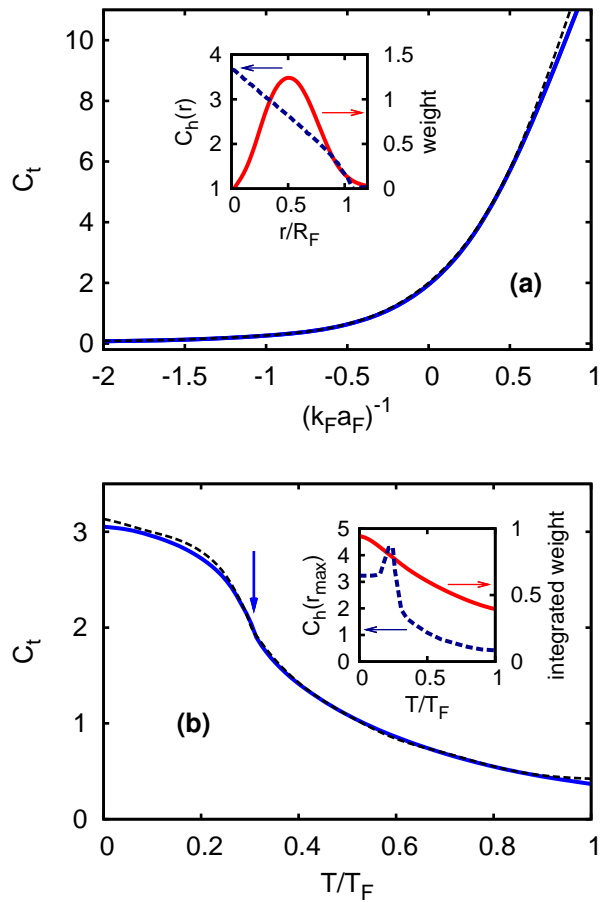}
\caption{The contact $\mathrm{C_{t}}$ obtained within the t-matrix approximation for the trapped case, is shown (full line) 
(a) at $T_{c}$ vs the coupling $(k_{F} a_{F})^{-1}$, and
(b) at unitarity vs $T/T_{F}$, and compared with the approximation (dashed line) that relies on the 
maximum of the radial weight function. 
[See the text for the meaning of the insets.]}
\end{center}
\end{figure} 



\begin{thebibliography}{99}

\bibitem{Tan-2008-I} S. Tan, Ann. Phys. {\bf 323}, 2952 (2008).

\bibitem{Tan-2008-II} S. Tan, Ann. Phys. {\bf 323}, 2971 (2008).

\bibitem{Braaten-2008} E. Braaten and L. Platter, Phys. Rev. Lett. {\bf 100}, 205301 (2008).

\bibitem{Castin-2009} F. Werner, L. Tarruell, and Y. Castin, Eur. Phys. J. B {\bf 68}, 401 (2009).

\bibitem{Leggett-2009} S. Zhang and A. J. Leggett, Phys. Rev. A {\bf 79}, 023601 (2009).

\bibitem{Combescot-2009} R. Combescot, F. Alzetto, and X. Leyronas, Phys. Rev. A {\bf 79}, 053640 (2009).

\bibitem{Castin-2010} F. Werner and Y. Castin, arXiv:1001.0774v1.

\bibitem{Jin-2010} J. T. Stewart, J. P. Gaebler, T. E. Drake, and D. S. Jin, Phys. Rev. Lett. {\bf 104}, 235301 (2010).

\bibitem{PPS-2009} P. Pieri, A. Perali, and G. C. Strinati, Nature Phys. {\bf 5}, 736 (2009), arXiv:0811.0770.

\bibitem{footnote-1} We set $\hbar = 1$ and $k_{B}=1$ throughout.

\bibitem{PPSC-2002} A. Perali, P. Pieri, G. C. Strinati, and C. Castellani, Phys. Rev. B {\bf 66}, 024510 (2002).

\bibitem{JILA-Cam-2009} J. P. Gaebler, J. T. Stewart, T. E. Drake, D. S. Jin, A. Perali, P. Pieri, and G. C. Strinati, 
                                          Nature Phys. (in press), and arXiv:1003.1147.
                                         
\bibitem{PS-2005} P. Pieri and G. C. Strinati, Phys. Rev. B {\bf 71}, 094520 (2005).        

\bibitem{Baym-2009} Z. Yu, G. M. Bruun, and G. Baym, Phys. Rev. {\bf 80}, 023615 (2009).  

\bibitem{Randeria-2009} W. Schneider and M. Randeria, Phys. Rev. A {\bf 81}, 021601 (2010).  

\bibitem{Braaten-2010} E. Braaten, D. Kang, and L. Platter, Phys. Rev. Lett. {\bf 104}, 223004 (2010).  

\bibitem{footnote-2} In weak coupling $k_{F} |a_{F}| \ll 1$, the zero-temperature mean-field gap $\Delta/E_{F} = (8/e^{2}) \, \mathrm{exp}\{  - \pi / (2 k_{F} |a_{F}|) \} $ associated with long-range order is sub-leading with respect to $\Delta_{\infty}$ associated with medium-range order.

\bibitem{PS-2003} P. Pieri and G. C. Strinati, Phys. Rev. Lett. {\bf 91}, 030401 (2003).

\bibitem{PPS-2004} The t-matrix approximation is extended to the superfluid case below $T_{c}$ following 
P. Pieri, L. Pisani, and G. C. Strinati, Phys. Rev. B {\bf 70}, 094508 (2004).

\bibitem{CGS-2006} R. Combescot, S. Giorgini, and S. Stringari, Eur. Lett. {\bf 75}, 695 (2006).

\bibitem{Belyakov-1961} V. A. Belyakov, Sov. Phys. JETP {\bf 13}, 850 (1961).

\bibitem{Chin-Julienne-2005} C. Chin and P. S. Julienne, Phys. Rev. A {\bf 71}, 012713 (2005).

\bibitem{PPS-2008} A. Perali, P. Pieri, and G. C. Strinati,  Phys. Rev. Lett. {\bf 100}, 010402 (2008).

\bibitem{footnote-3} The asymptotic behavior of transition matrix elements for large energy transfer depends on the short-range behavior of the two-body wave function [cf. A. R. P. Rau and U. Fano, Phys. Rev. {\bf 162}, 68 (1967)]. When final-state effects were taken into account, the RF signal would thus eventually acquire an $\omega^{-5/2}$ asymptotic tail 
\cite{Chin-Julienne-2005,PPS-2008}.

\bibitem{SM-PRL} See the Supplemental Material below reported for more details.

\end{thebibliography}
\end{document}